\documentclass[english,aps,reprint,nofootinbib]{revtex4-1}
\usepackage[T1]{fontenc}
\usepackage[utf8]{luainputenc}
\usepackage[letterpaper]{geometry}
\usepackage{float}
\usepackage{graphicx}
\usepackage{setspace}
\usepackage{color}
\usepackage{soul}

\makeatletter
 
 \@ifundefined{textcolor}{}
 {%
   \definecolor{BLACK}{gray}{0}
   \definecolor{WHITE}{gray}{1}
   \definecolor{RED}{rgb}{1,0,0}
   \definecolor{GREEN}{rgb}{0,1,0}
   \definecolor{BLUE}{rgb}{0,0,1}
   \definecolor{CYAN}{cmyk}{1,0,0,0}
   \definecolor{MAGENTA}{cmyk}{0,1,0,0}
   \definecolor{YELLOW}{cmyk}{0,0,1,0}
 }

\@ifundefined{showcaptionsetup}{}{%
 \PassOptionsToPackage{caption=false}{subfig}}
\usepackage{subfig}
\makeatother

\usepackage{babel}
 
\begin{document}

\title{Time Development of Early Social Networks: Link analysis and group
dynamics}

\author{Jesper Bruun}
\affiliation{University of Copenhagen, Department of Science Education}
\author{I.G. Bearden}

\affiliation{University of Copenhagen, Niels Bohr Institute}
\affiliation{University of Copenhagen, Department of Science Education}

\begin{abstract}
Empirical data on early network history are rare. Students
beginning their studies at a university with no or few prior connections
to each other offer a unique opportunity to investigate the formation
and early development of social networks. During a nine week introductory
physics course, first year physics students were asked to identify those with whom they communicated about problem solving in physics during
the preceding week. We use these students' self reports to produce time dependent student interaction networks. These networks have also been investigated to elucidate possible effects of gender and students' final course grade. Changes in the weekly number of
links are investigated to show that while roughly half of all links
change from week to week, students also reestablish a growing number
of links as they progress through their first weeks of study. To investigate
how students group, Infomap is used to establish groups. Further,
student group flow is examined using alluvial diagrams, showing that
many students jump between group each week., Finally, a segregation
measure is developed which shows that students structure themselves
according to gender and laboratory exercise groups and not according
to end-of-course grade. The results show the behavior of an early
social-educational network, and may have implications for theoretical
network models as well as for physics education.

\end{abstract}
\maketitle

\section{Introduction}

The formation and evolution of (social) networks has been modeled
by many researchers, who have have investigated theoretical models
of mechanisms for producing networks resembling empirical networks
\cite{barabasi1999emergence, Liben-Nowell2005, caldarelli2002scale, kim2005self, rosvall2006self}.
However, longitudinal network data is rare \cite{kossinets2006},
and it is difficult to obtain network data
from the time, $t_{0}$, at which the network begins to form. Here, we investigate longitudinal social network data from a time close to $t_{0}$.

Students beginning their university studies with few or no prior connections
to each other, are in a new situation, and will presumably make new
connections with other students as a natural part of their studies. Many of
them will also become socially involved, which also involves making
new connections to other students. As the students become both academically
and socially integrated they may change the ways in which they are connected,
and this might happen on a short time scale, perhaps daily or weekly.
Thus, investigating high resolution network data from such students
may offer insights as to how such networks form and how they evolve. 

If students beginning their studies do not know the other students,
we could expect them to try out many different possibilities for interaction.
Some of these interactions will be deemed worthwhile by the student,
and thus continue each week. Another would be that they
do not interact much at all, but work alone. Finally, one might expect that the high performing students would become central "nodes", due to word of mouth (i.e. as a student you would hear about student X
who is a high performing student, and then you would try to communicate
with him to understand the subject better). These early patterns of
interactions in social networks are largely unknown from an empirical
point of view. 

\begin{figure}
\includegraphics[width=1\columnwidth]{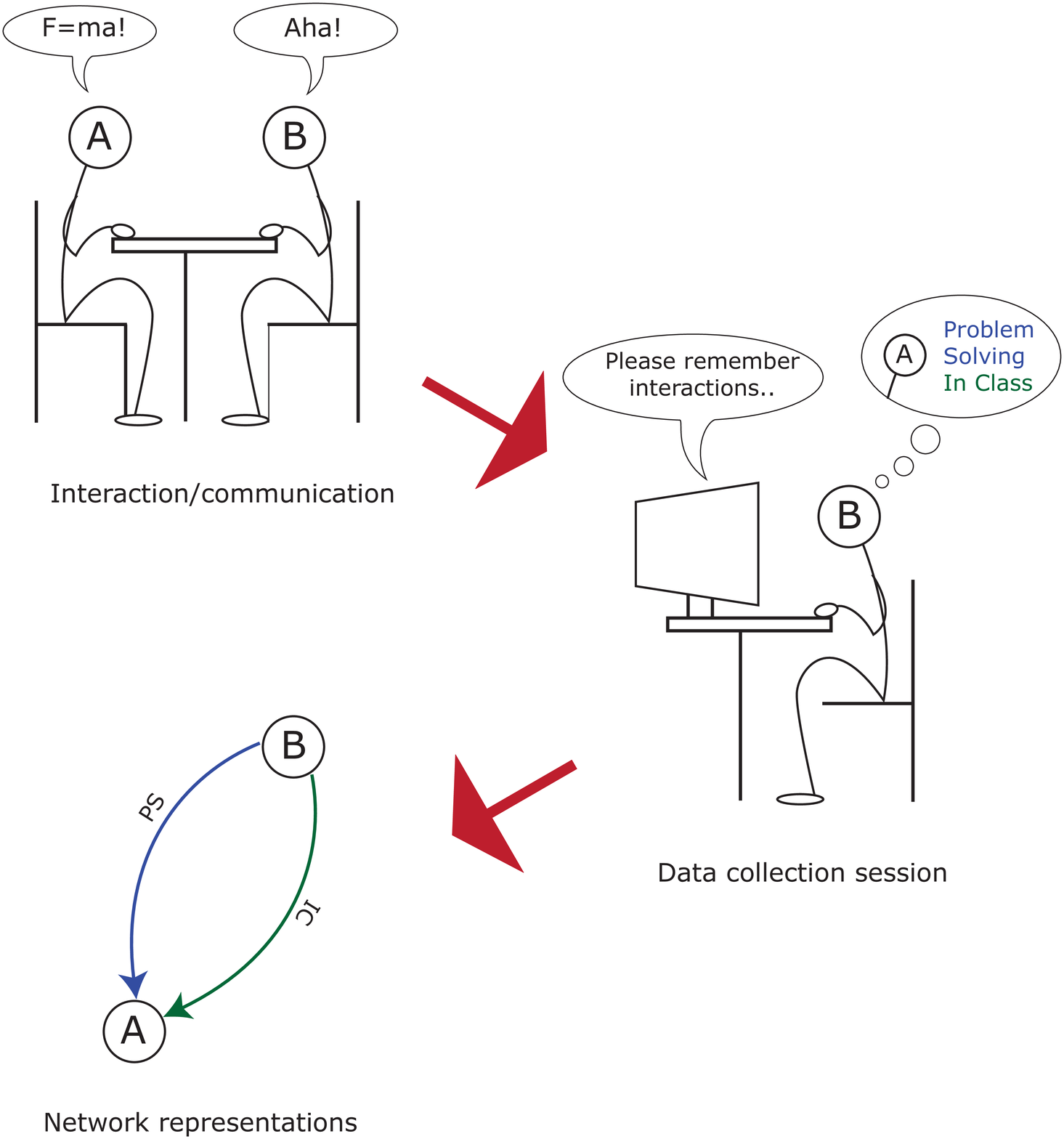}

 \caption{\label{fig:Students-interact-when}Students interacted
when studying and subsequently reported the interactions they remembered
via an online survey. Each remembered interaction becomes link in
a network. }

\end{figure}

This study investigates early interaction patterns among
approximately 170 students enrolled in an introductory mechanics course
at the University of Copenhagen. Students report which interactions
they remember in different categories (see Figure \ref{fig:Students-interact-when})
related to physics learning and social communication. Naming another
student naturally involves a direction (that is, A "names" B is A to B), so the network connections are directed.
Self-reported measures are notorious for being biased \cite{Liben-Nowell2005,liljeros2001web};
however, unlike potentially more objective measures, they can tell us what students are
interacting about. Also, using e-mails, phone calls, or digital proximity
as proxies for social ties, may be misleading. For
example, it has been found \cite{eagle2009} that people remember
their friends rather than remembering everyone they are near to as
measured by digital means. In contrast, asking students who they remember
having communicated with about some subject (in this case, physics
and social interactions), does indicate what the interaction was about.
In this work we try to minimize bias \cite{marsden11,liljeros2001web,pustejovsky2009question}
by only asking about remembered interactions and not asking students
to rank these relations in any way. In particular, we do not ask the students to judge the quality of the interaction in any way, althoughrecent work \cite{bruun2013talking} suggests that interactions remembered by students are more useful to them than are non-remembered interactions. 

To build an understanding of the processes underlying the formation and
evolution of social networks, researchers have related network measures
to non network node properties. That is, the effects on the network of individual actions by the nodes (in our case, students). For example, for university students
the the probability of making new social connections has been tied
to the number of classes taken together \cite{kossinets2006}. Also
in a twenty year long study, people with increasing body mass index
(BMI) tend to cluster together \cite{christakis2007spread}. Thus
relating the calculations we can perform on networks to the socially
relevant variables leads to knowledge of the social processes relevant
to network formation. 

One process occurring in social networks is the tendency for similar
nodes to connect or be directly connected, referred to as homophily \cite{kossinets2006}.
However, in many networks, nodes are seen to group together in clusters (or modules) Clusters are usually identified by community detection algorithms \cite{lancichinetti2009community}. Nodes belonging to the same cluster need not be similar or directly connected.
However, in these groups, information presumably flows more easily, and even if the probability
for a single node to associate with similar nodes is high, random connections to non-similar people will occur, thus diminishing
the similarity within the group.

Recently, the Infomap algorithm \cite{rosvall2008maps} has been used
to find clusters and inter-cluster structure in networks, employing
an information flow based perspective on grouping. Since the links
in this study represent student communication, this perspective seems
appropriate. Also, Infomap has been shown to perform well on directed
networks compared to other grouping algorithms \cite{lancichinetti2009community},
and changes of Infomap groupings of different networks with overlapping
nodes can easily be visualized with alluvial diagrams \cite{10.1371/journal.pone.0008694}.
Thus, we might expect Infomap to help find relevant groups of students interacting
with each other. 

Such groupings allow us to investigate how students structure themselves
during their first months at university. To quantify how students structure
themselves in groups we employ a measure of segregation
based on Kullback-Leibler divergence \cite{niven2005combinatorial}.
This measure is applied to each group to see how these groups
were segregated compared to the cohort's distribution of grade, gender, and class
number. Further, by giving each group's segregation a weight proportional
to the number of students in it, the segregation for the whole network
can be calculated. Thus, the segregation is a measure of how each group
and the whole network is structured according to a given attribute, compared
with the cohort's distribution. 

After explaining the background for data collection (Section \ref{sec:Background}),
we present empirical networks of self-reported interactions among the students in
 an introductory physics course at the University of Copenhagen
(Section \ref{sec:Resulting-networks}). Here we find that link
patterns change from week to week but that many links are reestablished
later on. Further, in Section \ref{sec:Group-flow} an alluvial diagram
shows how students jump between groups but that groups seem to stabilize
at the end of the measurement. Finally, the segregation measure is
developed in Section \ref{sec:Student-segregation}, and in Section
\ref{sec:Segregation-results} used with groups found with Infomap
to show that students do not structure themselves according to grade
but primarily according to their laboratory exercise groups, and somewhat
according to gender. This is followed by a conclusion in Section \ref{sec:Discussion-and-concluding}.

\section{Background\label{sec:Background}}

\subsection{Cohort and context}

Students were allotted time during the obligatory weekly laboratory
exercises to fill out online self-report surveys. Typically, students
would fill out the survey at some time during the lab exercise, although 
some chose to fill it out at home. They were encouraged to fill out
the survey at the beginning of a lab class, but some fitted in the
survey when a natural break came in their lab activities. Students
were told that their answers would be confidential and could not
be used by their instructors/lecturers to identify  individuals.
 Partcipation was not mandatory, although students were encouraged repeatedly to to take part in the study. 
The students in the course attend four hours of (non required) lecture per week. The students are assigned to sections, of which there are seven, and have the opportunity to attend four hours of \emph{problem solving sessions} .  Due to budgetary and space constraints, it is not possible to have the required laboratory exercises concurrently, so these are spread throughout the week. The same sections are also used in the introductory math course taken by these students.  Given this, students who attend all sessions on offer will spend at least 24 hours a week together, with roughly 15 hours of this spent in small (less than 30) sections of students. 

We can, therefore, assume that a student answered the survey at the same time
of week from one week to  another. That is, if student $A$ answers
the survey on a Tuesday afternoon one week, chances are that student
$A$ answers the survey on the following Tuesday again. However, students
were allowed to switch lab exercise hours, if for some reason they
were not able to make it to the scheduled one.  Typically, 3 or 4 students per week attended another section. Thus, there is some
fuzziness with regards to when student answers are recorded. 

The measurements were done during a course in introductory mechanics
and special relativity at a large Danish University. Students are primarily
ethnic Danes. The majority (roughly 85\%) of students are physics majors who have
just started their studies. Some major in other disciplines (for example
mathematics), but are allowed to choose this physics course as part
of their study plans.

\subsection{Description of survey and data collection}

The online survey was divided into two parts each week; an academic part
and a social part. The academic part consisted of 9 interaction categories,
while the social part consisted of 3 interaction categories. The categories
were developed through a mixed methods pilot project prior to data
collection \cite{bruun2012}, and in this study we only examine the
category pertaining to \emph{communication concerning problem solving}.
A weekly format was chosen based on \cite{eagle2009} who found the
greatest correspondence between self reported networks and digitally
measured proximity networks if the interactions were reported within
a week. While more finely grained temporal data would be interesting, conversations with a number of students indicate that asking for answers more often than once per week would lead to survey fatigue and significantly lower participation. 

Students were given a login to a learning management system, where
they could take the survey each week. For each interaction category,
students marked each of the students they remembered having had interactions
with. Names of possible students (all students enrolled in the course)
were given in a roster\cite{marsden11}. The researcher was present
throughout most data collection sessions, and students were invited
to ask if they had doubts on how to answer the survey. The researcher
emphasized repeatedly that they should mention only the people they
remembered, that their answers where anonymous, and that there was
no implicit ranking of their friends.

\section{Resulting networks \label{sec:Resulting-networks}}

\begin{figure}
\includegraphics[height=1\textwidth]{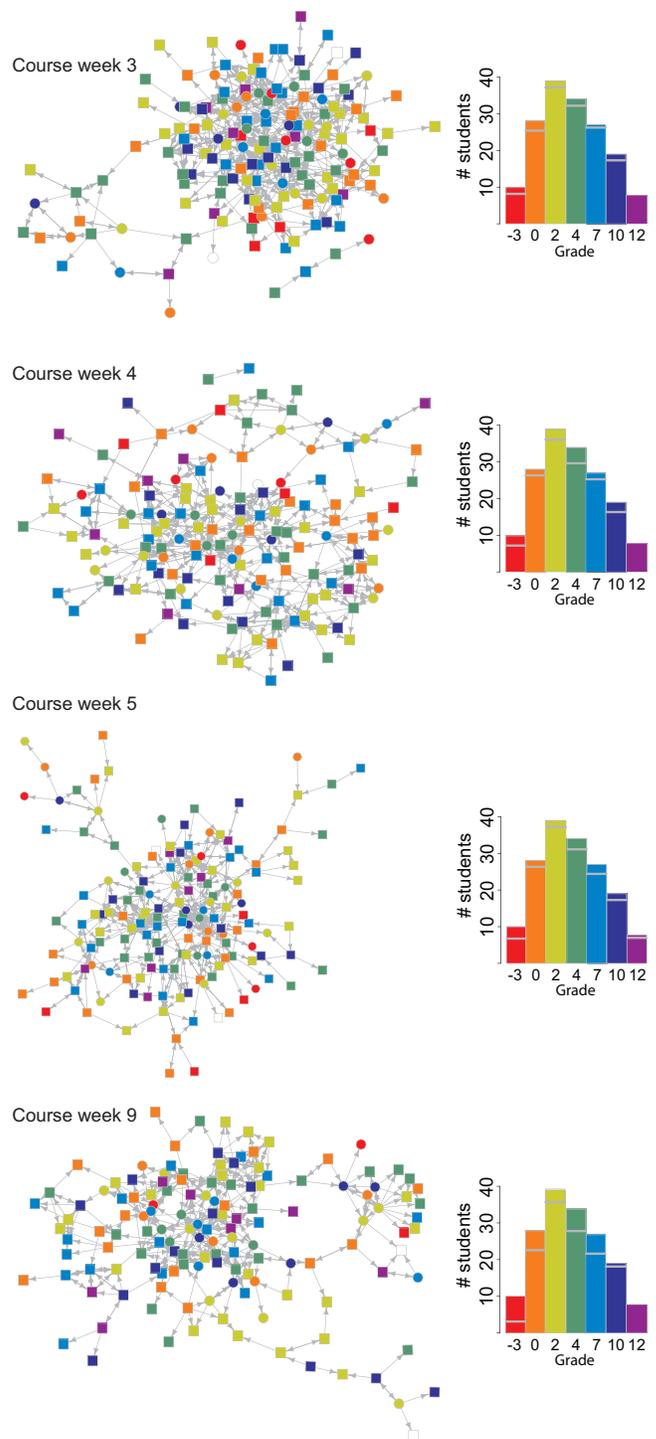}

\caption{ \label{fig:Four-networks-from}Four networks from different
weeks. The density $\rho\approx0.02$ for each network. Color codes
represent end-of-course grades: -3 and 0 are failed. The histograms
show how many students with a particular grade are present in the
network compared with the total number of students getting a grade.
Females are represented by circles, males by squares. The total number
of of nodes in the networks are 161, 152, 154, and 139 respectively. }
\end{figure}

The introductory course has a duration of nine weeks, but there are
only seven networks, four of which are displayed in Figure \ref{fig:Four-networks-from}.
Due to initial confusion about how to respond, many students did not
answer the first two surveys, so the first two weeks' data were combined. Further, due to a technical error, course week
6 data were not recorded. The networks do not seem to indicate much
segregation according to  the final grade earned in the course. The non-passing students (red
and orange nodes) seem to move from being well integrated in the network
to  the periphery in week 5 and most red nodes are gone
in week 9. This may be explained by the structure of the course,
where students are continuously solving problems and getting grades.
Thus, students who do poorly might simply have left the course. On
the other hand, they might be pursuing the course  without
participating in the survey or being named by others. 

Turning to the structure of the networks, they evolve from a
compact to a more stringy nature. This may signify students' tendency
to form groups which are connected by bridges \cite{wasserman1994social,scott2011sage}
as proposed by social network analysis. However, there may also be
effects of survey participation since the number of students naming
at least one other student is  124, 115, 98, and 83 for
the four networks. This decline in participation could be due to student fatigue with respect
to the survey \cite{pustejovsky2009question}, an increased workload
on the students towards the end of the course,  student drop out, 
or to a combination of the three. 

\begin{figure}[H]
\subfloat{\includegraphics[width=0.75\columnwidth]{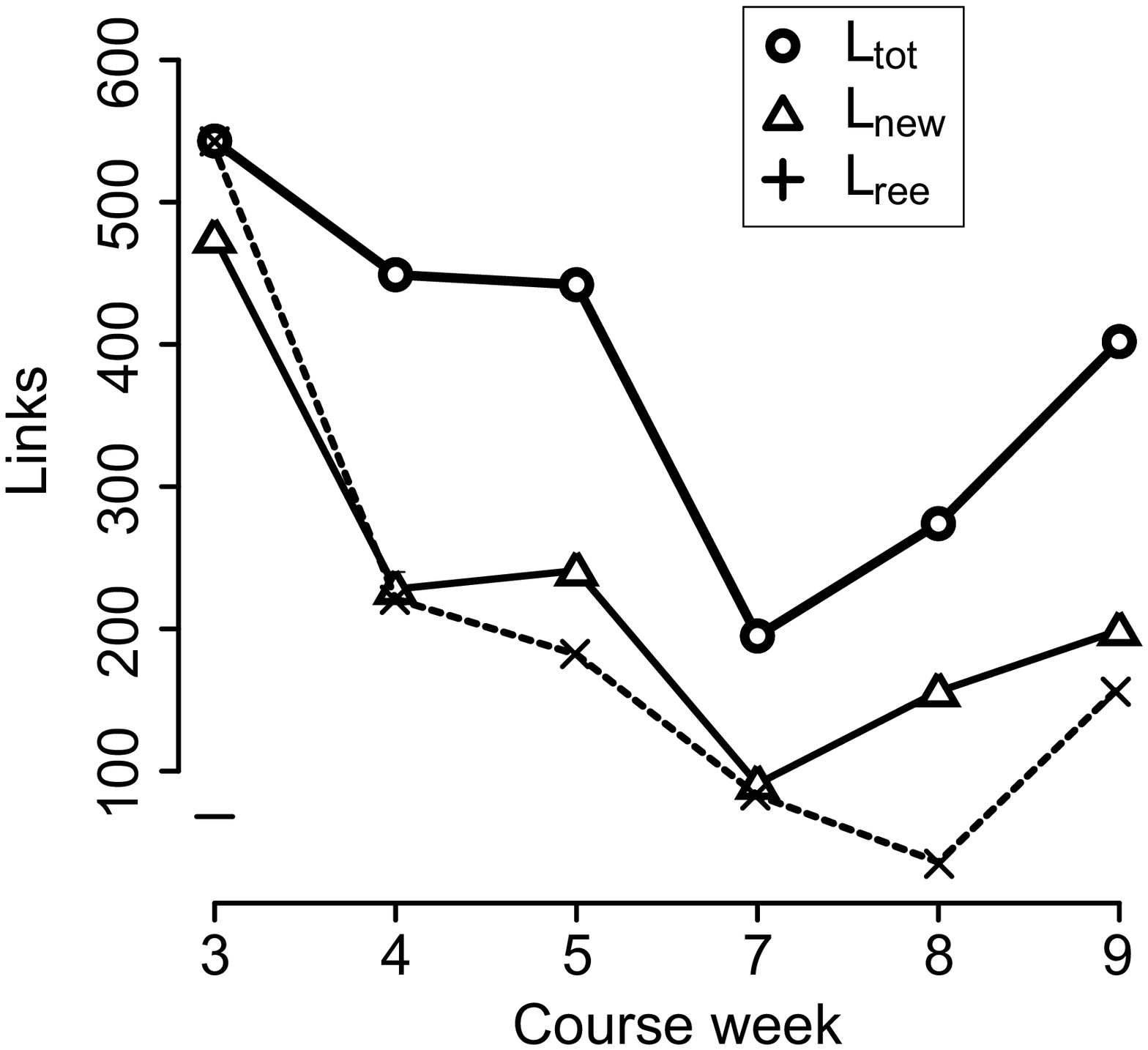}}

\subfloat{\includegraphics[width=0.75\columnwidth]{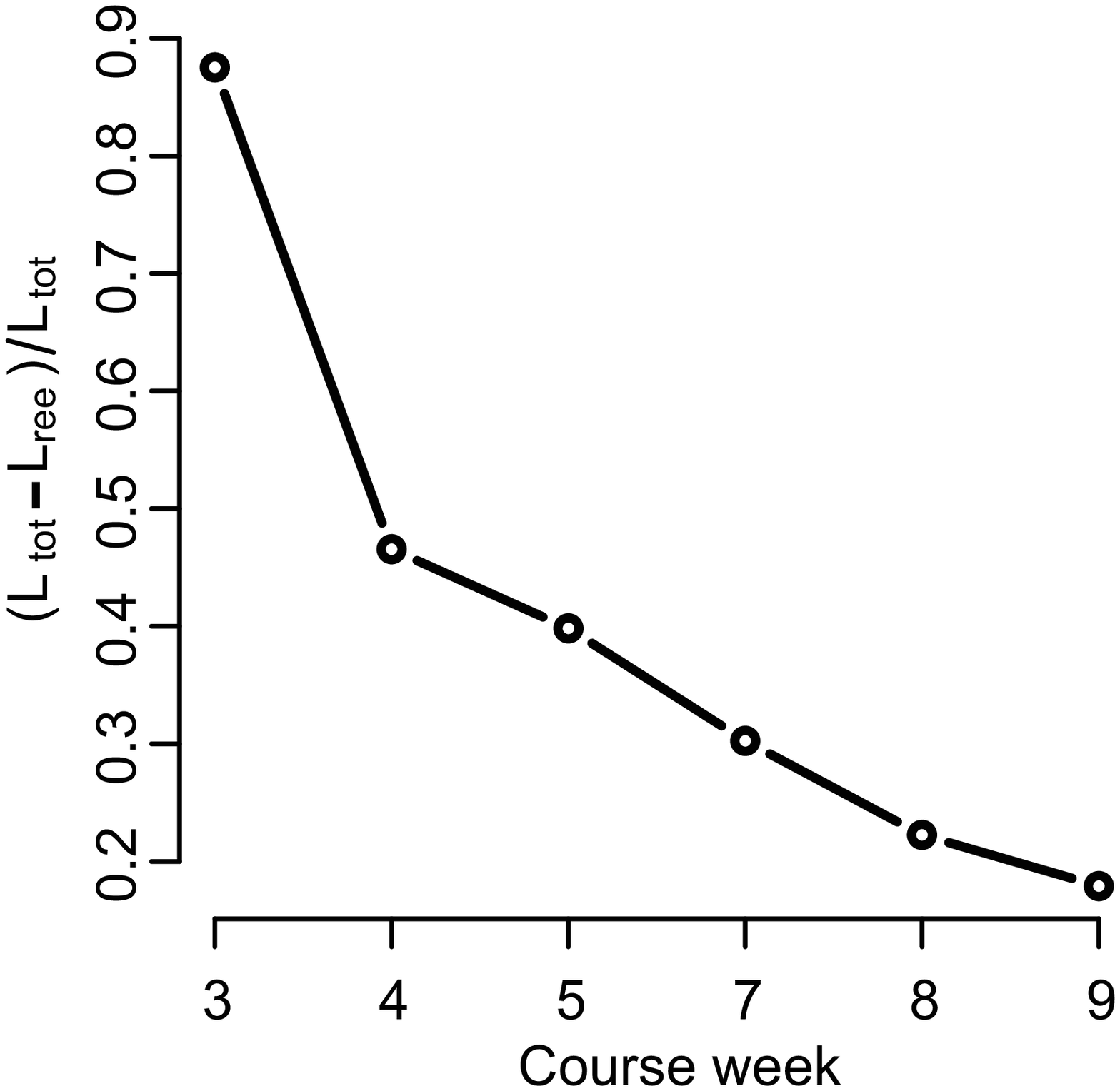}}

\caption{\label{fig:Link-patterns-weeks} (a) Number of total,
new, and reestablished links. New links, $L_{new}$, calculated based
on the preceding week. $L_{ree}$ is the number of links in a given
week which are also present in at least one of the preceding weeks.
$L_{tot}=160$ for week 2 (not shown). (b) The fraction of completely
new links relative to the total number of links, $\frac{L_{tot}-L_{ree}}{L_{tot}}$
, seems to decrease exponentially.}
\end{figure}

Figure \ref{fig:Link-patterns-weeks} (a) show how the total number
of links, $L_{tot}$ change from week to week. The dip in course week
7 corresponds to the traditional Fall Break in most Danish educational
institutions. However, it is peculiar, since this is an intensive
course with no scheduled fall break. However, this would explain both
the dip and the slow recovery in course week 8: In week 7 a larger
number of students would be absent thus not answering the survey.
In week 8 few people would list having had physics interactions with
these students. 

There are a considerable number of new links, $L_{new}$, each week
compared with the preceding week. Roughly half of the links each week
are new compared to the preceding week. However, the number of re-established
links, $L_{ree}$, comprise a larger and larger fraction of the total.
For a given week, the number of re-established links is the number of links
in the network which are present in at least one of the preceding weeks' networks. 
Together, the variations in $L_{new}$ and $L_{ree}$ may be used
to form a hypothesis about how bonds are created during the early
stages of this particular student network's history: students try
working with a lot of different collaborators. As they do this, they
find out who they want to work with and return to them again. This
is supported by Figure \ref{fig:Link-patterns-weeks} (b), where the
fraction of completely new links, $\frac{L_{tot}-L_{ree}}{L_{tot}}$,
is shown to decrease over time. The number of unique links for
all weeks is 1214, which is roughly 5\% of the total number of possible
links ($L_{possible}=N(N-1)$ in a directed network with 140-160 nodes ($L_{possible}> 19500$). This implies that the decrease
in completely new links is not due to a saturation of the network
links.

\section{Group flow\label{sec:Group-flow}}

\begin{figure*}
\includegraphics[width=1\textwidth]{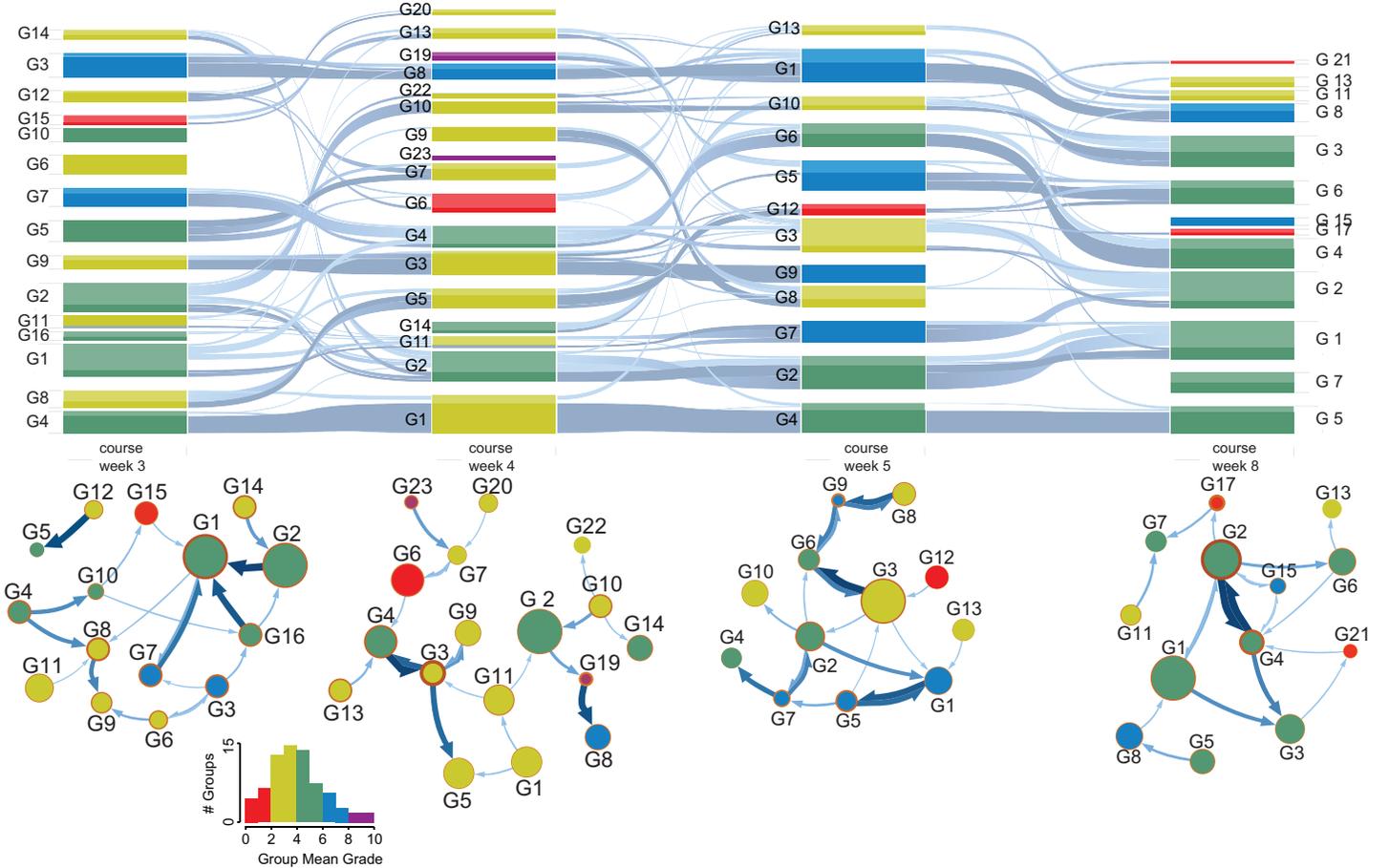}

\caption{\label{fig:Alluvial-diagram-fourNetworks}Top: Alluvial
diagram for groupings of the four networks displayed in Figure \ref{fig:Four-networks-from}.
Each column in the diagram account for approximately 80\% of the total
PageRank corresponding to around 70\% of the students in the network
for the corresponding course week. The height of a block representing
a group is proportional to the accumulated PageRank of the group,
and the lighter colors in each group indicate how much PageRank is
insignificantly clustered. Finally, the thickness of the gray streamlines
between groups in different weeks indicate the initial and final PageRank
the nodes making the transition from one group to another. Bottom:
Flow maps of the same groups shown in the alluvial diagram. Node sizes
are proportional to the number of students in the group. Arrows are
proportional to the information flow between groups as calculated
by Infomap. The total number of groups each week is 28,28,22,22, respectively.
Groups are color coded according to their mean grade. The histogram
shows the distribution of grades for all groups in all networks.}
\end{figure*}

Figure \ref{fig:Alluvial-diagram-fourNetworks} shows the alluvial
diagram \cite{10.1371/journal.pone.0008694} for student groups in
the four networks displayed in Figure \ref{fig:Four-networks-from}.
The height of each is box proportional to the accumulated PageRank \cite{brin1998anatomy}
of the group. The color of the boxes mark what range the group mean
grade falls in. Though there does not seem to be a connection between
group mean grade and accumulated PageRank, the group mean grades seem
to become more homogeneously distributed among the groups in the diagram
in course week 9 compared with the preceding weeks.

The stream lines between each column indicates shifts in PageRank
from one week to the other. Some groups seem stable throughout the
course, but many changes happen between weeks. However, there are
fewer stream lines between week 5 and 9 (38) than between the other
weeks (46 and 51 respectively), which indicates that groups seem to
stabilize somewhat over time. However, on a week to week basis groups
seem to change a lot, especially in the beginning of the course.

Most boxes have a light and dark shade. The light shade indicates
the accumulated PageRank of the students which are not significantly
(90\% confidence) attached to the group in question as found by a
bootstrapping procedure \cite{10.1371/journal.pone.0008694}. This
indicates that the students, which these light shaded boxes represent
could not be reliably assigned to the group in question in a "bootstrap
world" of resampled network. 

Network maps are also shown in Figure \ref{fig:Alluvial-diagram-fourNetworks}.
These map show groups of students as nodes with sizes proportional
to the number of students in each group. They are again color coded
according to the mean grade of the group, showing that there does
not seem to be a clear cut relation between the number of people in
the group and the group mean grade. 

The arrows indicate probability flow, which we can relate to how many
students in one group name students in another group. Since these
namings indicate communication about problem solving, these arrows
might indicate which groups are important for how problem solving
knowledge is spread in the network. As such, they might provide an indication of which students need and which students can give help in an introductory physics course. 

PageRank is an interesting measure for many networks, but here we
can note that it would probably be more interesting if the stream
lines and height of boxes represented actual students. Then we would
be able to see more clearly which students move, what their attributes
are, and how this affects group composition. We would also be able
to see clearly, which students are difficult to assign into groups.
Thus, in future work, it might be beneficial to change the interpretation
of sizes in the alluvial diagram.

\section{Student segregation\label{sec:Student-segregation}}
To quantify how students with different attributes segregate, assume
an attribute with $n$ possible values. For example, gender would have
$n=2$. A grouping algorithm now creates a grouping $M$ with $N_{g}$
different groups. Choosing a node at random in the network, the probability
of choosing a female is $q_{female}=\frac{N_{female}}{N}$. However, groups
potentially have other distributions of males and females, so in the
$i'th$ group, $p_{female}^{i}=\frac{N_{female}^{i}}{m_{students}^{i}}$.
In an information theoretical approach \cite{niven2005combinatorial},
the \emph{surprisal}
\footnote{This is actually our cross-surprisal, that is the information gained
relative to the information we had prior to investigating the grouping.%
} is $\delta_{female}^{i}=log_{2}(\frac{p_{female}^{i}}{q_{female}})$. For
the whole group, our expectation of surprisal for the $i'th$ group
is $D^{i}=p_{female}^{i}log_{2}(\frac{p_{female}^{i}}{q_{female}})+p_{male}^{i}log_{2}(\frac{p_{male}^{i}}{q_{male}})$.
This is an instance of Kullback-Leibler divergence \cite{niven2005combinatorial}.
Treating the groups as independent sub-systems, the total weighted
segregation for this particular grouping $M$ is $D(M)=\sum_{i}\frac{m_{students}^{i}}{N}D^{i}$. 

In general, an attribute may have more possible values. If a student
may only take one of these values, probability distributions $\{p_{ij}\}$
and $\{q_{j}\}$ can be defined so that:

\begin{equation}
D_{seg}(M)=\frac{1}{N_{s}}\sum_{i\in M}m_{i}(\sum_{j=1}^{n}p_{ij}\log_{2}(\frac{p_{ij}}{q_{j}}))
\end{equation}

where $N_{s}=\sum m_{i}$ is the total number of students in the grouping.
The range of $D_{mix}(M)$ can be estimated as follows: If $p_{ij}=q_{j}$for
all $j$ in all groups $m_{i}$, $D_{seg}^{min}(M)=0$. For perfect
segregation, where for each group $p_{j}=\delta_{lj}$ for the $l'th$
of the $n$ categories the segregation is $D_{seg}^{max}=-\sum_{j=1}^{n}q_{j}log_{2}(q_{j})$%
\footnote{$D_{nomix}(M)=\frac{1}{N_{s}}\sum_{i\in M}\sum_{l=1}^{n}m_{i}^{l}\log(q_{l}^{-1})$.
Adding groups with same $l$ yields $D_{nomix}(M)=\frac{1}{N_{s}}\sum_{l=1}^{n}m_{l}\log(q_{l}^{-1})$.
Finally use $q_{l}=\frac{m_{l}}{N_{s}}$. %
}. 

Here, the segregations according to different attributes are calculated:
Gender ($n=2$), grade ($n=7$), and lab class ($n=7$). To see how
different the segregations are from a random distribution of gender,
grade, or lab class, the attributes are resampled, while keeping the
module structure $M$. It corresponds to changing the distribution,
$\{p_{ij}\}$, in each group at random, while keeping the prior distribution,
$\{q_{i}\}$. This resampling is done a number of times (here, $10^{4})$
and each time $D_{seg}^{r}(M)$, is calculated. Finally, the Z-score,
$Z=\frac{D_{seg}(M)-\left\langle D_{seg}^{r}(M)\right\rangle }{\sigma_{r}}$,
is calculated. Thus, the results will show the deviation from random
variations. 

\section{Segregation results\label{sec:Segregation-results}}

\begin{figure}
\includegraphics[width=0.75\columnwidth]{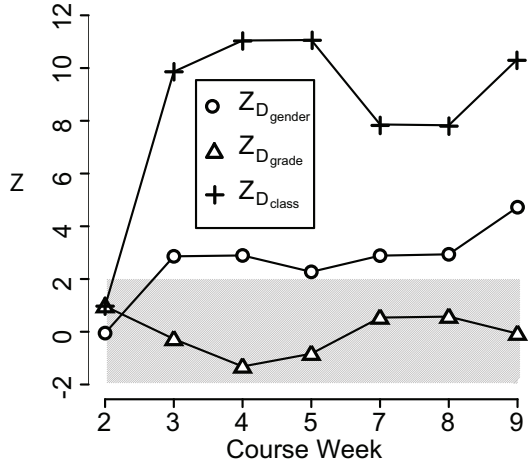}

\caption{ \label{fig:Segregation-z-scores} Segregation Z-scores
for gender, grade, and lab class for each week. The shaded area indicates
the non significant region.}
\end{figure}

The results of the calculations for the whole network segregation
from week to week during the course are shown in Figure\ref{fig:Segregation-z-scores}
(a). The expected distribution $\left\{ q_{i}\right\} $ is calculated
from the student present in the network. The first week shows neither significant
segregation nor non-segregation. During the following weeks, students
segregate significantly according to lab classes and to a lesser degree
according to gender. 

While there is significant segregation according to gender and lab
class , students do not segregate or mix near perfectly. If students
segregated perfectly, calculations show that the Z-scores would be
around 20 for gender and around 40 for grade and lab class. If they
did not segregate at all, that is if $D_{seg}=0$ corresponding to
perfect mixing, the Z-scores would be around -2 for gender and -4
for grade and lab class. Thus, groups do not consist for example of
students from only one lab class but of clusters of students from
different classes.

\section{Conclusions\label{sec:Discussion-and-concluding}}

This study examined the early stages of network formation based on
student reports of who they remember having communicated with about
problem solving in physics. Seven networks made from weekly reports
these types of communication in an introductory physics course were
analyzed. In these networks less and less low performing students
are represented, but the remaining students do not to segregate according
to end-of-course grade. 

The link analysis showed that roughly half of the links in a week
were new compared with the preceding week. However, as the weeks go
by, students communicate with former communication partners, which
is indicated by the relative decrease in total links. Group flow patterns
were were examined. Stream lines in alluvial diagrams show that while
some groups seem stable throughout the different weeks, students seem
to flow extensively between groups. Also, the alluvial diagram revealed
that many students could not be significantly clustered together in
groups. Finally, student segregation was analyzed using a divergence
measure, called the segregation, which was applied to groups found
with the Infomap grouping algorithm. This analysis showed that students
segregate significantly compared to according to lab class number
and to lesser extent gender, but not according to grade. It was also
shown how individual groups found by Infomap could be analyzed. 

The overall picture painted by these analyses is that students
try out many different possibilities for collaboration in the first weeks
but gradually settle to communicate with the same people. Some find
study partners based on lab/problem solving classes and to some extent
gender, with whom  they continuously collaborate or reconnect with
during the course. However, it is generally difficult for infomap
to find assign these students to only one group. Further research
could use the results from the link analysis to constrain models of
network development. Another direction would be to further investigate
Infomap groupings. Since Infomap yields information about which alternative
groups it could find using bootstrap worlds, these groups could also
be investigated with the segregation measure. As a final note, these
results also have value for physics education research. 

\begin{acknowledgments}

We gratefully acknowledge the physics students who answered the survey. Without their help, there would be no network data. We also thank Professor
Kim Sneppen for useful suggestions and discussions. 

\end{acknowledgments}

\bibliographystyle{apsrev4-1}
\bibliography{/Users/jesperbruun/Dropbox/IND/Litteratur/jbReferences}

\end{document}